\documentstyle[twoside,fleqn,espcrc2my]{article}

\newcommand{\nc}{\newcommand}
\newcommand{\rnc}{\renewcommand}

\def\bo{\raisebox{-.4ex}{\large$\Box$}}                 
\def\={\ =\ }
\def\+{\ +\ }
\def\-{\ -\ }

\nc{\id}{{\bf 1}}
\nc{\Tr}{{\rm Tr\,}}
\nc{\rome}{{\rm Roma}}
\nc{\ie}{{\em i.e.}}
\nc{\eg}{{\em e.g.}}
\nc{\etal}{{\em et al.}}
\nc{\calF}{{\cal F}}
\nc{\calL}{{\cal L}}
\nc{\calS}{{\cal S}}
\nc{\kapbar}{\bar{\kappa}}
\nc{\uidot}{\dot u}
\nc{\uiidot}{\ddot u}
\nc{\uiiidot}{\stackrel{\ldots}{u}}
\nc{\psii}{\psi^{(1)}}
\nc{\psibar}{\overline\psi}
\nc{\chibar}{\overline\chi}
\nc{\psiibar}{\overline\psi^{(1)}}
\nc{\psibari}{\overline\psi^{(1)}}

\rnc{\topfraction}{1.0}
\rnc{\bottomfraction}{1.0}
\rnc{\textfraction}{0.0}

\nc{\qq}{\P}
\nc{\qqc}{[\P]}

\nc{\rng}{\rangle}
\nc{\lng}{\langle}
\nc{\rcite}{ref.\ \cite}
\nc{\ba}{\begin{array}}
\nc{\ea}{\end{array}}
\nc{\lb}{\left(}
\nc{\rb}{\right)}
\nc{\qrt}{\frac{1}{4}}

\nc{\al}{\alpha}
\nc{\bt}{\beta}
\nc{\gm}{\gamma}
\nc{\dl}{\delta}
\nc{\ep}{\epsilon}
\nc{\varep}{\varepsilon}
\nc{\zt}{\zeta}
\nc{\et}{\eta}
\nc{\kp}{\kappa}
\nc{\lm}{\lambda}
\nc{\rh}{\rho}
\nc{\sg}{\sigma}
\nc{\ta}{\tau}
\nc{\ph}{\phi}
\nc{\vr}{\varphi}
\nc{\ch}{\chi}
\nc{\ps}{\psi}
\nc{\om}{\omega}
\nc{\noi}{\noindent}
\nc{\half}{{\textstyle \frac{1}{2}}}
\nc{\rr}[1]{$^{#1}$}
\nc{\rf}[1]{(\ref{#1})}
\nc{\rfs}[2]{(\ref{#1},\ref{#2})}
\nc{\smgr}{\stackrel{\textstyle <}{>}}
\nc{\grsm}{\stackrel{\textstyle >}{<}}
\nc{\aleq}{\mbox{}_{\textstyle \sim}^{\textstyle < }}
\nc{\ageq}{\mbox{}_{\textstyle \sim}^{\textstyle > }}
\nc{\ra}{\rightarrow}
\nc{\lra}{\leftrightarrow}
\nc{\be}{\begin{equation}}
\nc{\ee}{\end{equation}}
\nc{\bea}{\begin{eqnarray}}
\nc{\eea}{\end{eqnarray}}
\nc{\eqrf}{eq.\ \rf}
\nc{\erf}{{\rm erf}}

\nc{\ap}[1]{Ann.\ Phys.~#1\ }
\nc{\app}[1]{Acta Physica Polonica~#1\ }
\nc{\arnps}[1]{Ann.\ Rev.\ Nucl.\ Part.\ Sci.~#1\ }
\nc{\cmp}[1]{Commun.\ Math.\ Phys.~#1\ }
\nc{\cpc}[1]{Comput.\ Phys.\ Commun.~#1\ }
\nc{\jetp}[1]{Soviet Physics JETP~#1\ }
\nc{\jpa}[1]{J.\ Phys.\ A~#1\ } 
\nc{\jpg}[1]{J.\ Phys.\ G~#1\ } 
\nc{\mpla}[1]{Mod.\ Phys.\ Lett.~A#1\ }
\nc{\npa}[1]{Nucl.\ Phys.~A#1\ }
\nc{\npb}[1]{Nucl.\ Phys.~B#1\ }
\nc{\nproc}[1]{Nucl.\ Phys.~B (Proc.\ Suppl.)~#1\ }
\nc{\pla}[1]{Phys.\ Lett.~#1A\ }
\nc{\plb}[1]{Phys.\ Lett.~#1B\ }
\nc{\pr}[1]{Phys.\ Rep.~#1\ }
\nc{\pra}[1]{Phys.\ Rev.~A#1\ }
\nc{\prb}[1]{Phys.\ Rev.~B#1\ }
\nc{\prc}[1]{Phys.\ Rev.~C#1\ }
\nc{\prd}[1]{Phys.\ Rev.~D#1\ }
\nc{\pre}[1]{Phys.\ Rev.~E#1\ }
\nc{\prep}[1]{Phys.\ Rep.~#1\ }
\nc{\prev}[1]{Phys.\ Rev.~#1\ }
\nc{\prl}[1]{Phys.\ Rev.\ Lett.~#1\ }
\nc{\procroy}[1]{Proc.\ Roy.\ Soc.~#1\ }
\nc{\ptp}[1]{Prog.\ Theor.\ Phys.~#1\ }
\nc{\rmp}[1]{Rev.\ Mod.\ Phys.~#1\ }
\nc{\rpp}[1]{Rep.\ Prog.\ Phys.~#1\ }
\nc{\sjnp}[1]{Sov.\ J.\ Nucl.\ Phys.~#1\ }

\newcommand{\AmS}{{\protect\the\textfont2
  A\kern-.1667em\lower.5ex\hbox{M}\kern-.125emS}}

\title{Laplacian Abelian Projection\thanks{To appear in the proceedings of
 ``Lattice 96'', St.\ Louis, Missouri, USA, June 4--8, 1996.\protect\\
 Work supported by EC contract ERBCHBICT941067 and by DGICYT project AEN 94-218.}
\hfill {\sf DFTUZ/96/17}}

\author{A.J. van der Sijs\address{Departamento de F\'\i sica Te\'orica,
           Universidad de Zaragoza, Facultad de Ciencias, 
           50009 Zaragoza, Spain\\ {\protect\sf [arjan@melkweg.unizar.es]}}}

\begin{document}

\thispagestyle{empty}

\begin{abstract}
A new partial gauge fixing condition for the abelian projection is introduced.
It is based on the lowest-lying eigenvector of a covariant Laplacian operator.
This gauge is smooth and free of lattice Gribov copies.
These properties are important for an unambiguous computation of the
abelian projected gauge field configuration.
\end{abstract}

\maketitle

\renewcommand{\thefootnote}{\fnsymbol{footnote}}

\section{INTRODUCTION}

The abelian projection \cite{thooft} remains the most popular framework for
lattice Monte Carlo studies of monopoles and confinement.
There are two uncomfortable aspects to the partial gauge fixing involved,
though.
One is the apparent preference for one particular gauge, the Maximally Abelian
Gauge (MAG) \cite{kron2}, the other is the presence of lattice Gribov copies
in this gauge.

The situation with respect to the first point can be loosely
summarized as follows.
The MAG ``works well'',  the temporal (``Polyakov'') gauge
``works sometimes'', and the other gauges ``do not work''.\footnote[7]{However,
for some new gauge conditions, see \rcite{suzuki96}.}\@
(See the reviews
\cite{suzukiosaka,osaka,greensite}
for results and interpretations.)\@
This may have its origin in the different smoothness 
properties of these gauges.
The MAG is a smooth gauge, the Polyakov gauge is smooth in the time direction
only, while the other gauges may not be smooth enough.
Smoothness means that the link matrices are relatively close to unity,
which is important for interpreting them in terms of continuum gauge
fields, and hence for extracting their abelian part.

Turning attention to the apparently preferred MAG then,
one is confronted with the second issue:
In practice, implementation of the MAG is complicated by the
presence of lattice Gribov copies, corresponding to different
local minima of the gauge fixing functional.  
Unambiguous computation of the gauge fixed configuration is impossible,
and certain quantities such as the monopole density 
are fairly sensitive to this ambiguity \cite{born,hioki}.
One would like to have a smooth gauge without lattice Gribov problem.

An analogous problem arises in conventional Landau gauge fixing.  Its
lattice implementation is smooth but there are lattice Gribov copies.
To circumvent this problem, Vink and Wiese introduced ``Laplacian gauge 
fixing'' \cite{vinkwiese}, which shares the smoothness properties of 
the Landau gauge but avoids lattice Gribov copies.  
In this proposal, the gauge transformation matrices are determined in
terms of the lowest-lying eigenvectors of the covariant Laplacian in
the background of the given gauge field configuration.
Recently this method was studied in practice \cite{vink}, and its 
perturbative formulation was given \cite{vanbaal}.

Here I take this idea over to the abelian projection, constructing a
procedure for partial gauge fixing which is smooth and free from
ambiguities.
Section 2 introduces the gauge, discusses its expected merits and
briefly describes its perturbative continuum formulation.
Numerical results are presented in section 3.

\section{LAPLACIAN ABELIAN GAUGE}
To introduce the Laplacian Abelian Gauge (LAG) it is convenient to start
from the MAG in its spin-model formulation \cite{thesis}.
I will limit myself to the SU(2) case here.

The Maximally Abelian Gauge \cite{kron2} for a link configuration 
$\{U_{\mu,x}\}$ is defined as the 
configuration $\{ \bar U_{\mu,x} \! = \bar\Omega_x U_{\mu,x} 
\bar\Omega^+_{x+\hat\mu}\}$ where $\{ \bar\Omega_x \}$
minimizes the functional
\be
\tilde\calS_U(\Omega) \= \sum_{x,\mu} \left\{ 1 -
 \half \Tr \left[\sigma_3 U_{\mu,x}^{(\Omega)}
            \sigma_3 U_{\mu,x}^{(\Omega)\,+} \right] \right\} ,
\label{S1}
\ee
which can be written in the form
\bea
\calS_U(\phi) &=& \sum_{x,\mu}
 \left\{ 1 - \half \Tr [ \Phi_x U_{\mu,x} \Phi_{x+\hat\mu} U^+_{\mu,x}]\right\} 
\label{S2a}  \\
 &=& \sum_{x,\mu}
 \{ 1 - \sum_{a,b} \phi^a_x R^{ab}_{\mu,x} \phi^b_{x+\hat\mu} \} 
 , \label{S2b}
\eea
with the definitions
\bea
\Phi_x &=& \Omega^+_x \sigma_3 \Omega_x
       \= \sum_{a=1}^3 \phi^a_x \sigma_a
 , \label{phi} \\
R^{ab}_{\mu,x} &=& \half \Tr [ \sigma_a U_{\mu,x} \sigma_b U^+_{\mu,x} ]
 . \label{Rab}
\eea
$R_{\mu,x}$ is the link matrix in the adjoint representation and
$\phi_x \in SU(2)/U(1) \simeq S^2$ is a three-vector of unit length,
neutral under abelian gauge transformations, as expected.
The gauge fixing functional \rf{S2b} is nothing but the latticized
covariant kinetic action $\int \half (D_\mu \phi)^2$
of a spin field $\phi$ in the background of the given gauge field.
The spin configuration which minimizes this action determines the MAG.
Finding this absolute minimum is difficult because of the length-one 
constraints on the individual spins.
The usual iterative local algorithms often get stuck in a local minimum.
This means that the result of the gauge fixing depends on the particular
algorithm used and on the starting point on the gauge orbit:
the gauge fixing procedure is ambiguous.

The idea of Laplacian gauge fixing is to:

1.\ Minimize $\calS_U(\phi)$ without taking into account the constraints
$\| \phi_x \| = 1$.
This amounts to finding the eigenvector $\bar\phi^a_x$
belonging to the lowest eigenvalue $\lambda$
of the covariant Laplacian $-\bo^{a\,b}_{x\,y}(R)$.

2.\ Write the solution $\bar\phi^a_x$ as
$\bar\phi^a_x = \rho_x \hat\phi^a_x$ and take $\hat\phi^a_x$ for the gauge
transformation: $\hat\Phi_x = \sum \hat\phi^a_x \sigma_a
= \bar\Omega^+_x \sigma_3 \bar\Omega_x$
(of course $\bar\Omega_x$ is determined up to the residual
U(1) freedom only).

This procedure is unambiguous because the computation of eigenvectors
can be done to the precision required, and because the procedure is gauge
covariant by construction: under a gauge transformation $V$ of the 
starting configuration
$U$, the Laplacian operator $-\bo(R(U))$ and its eigenvectors transform
accordingly, such that the gauge fixed configuration $\bar U$ is 
unchanged
(again, up to residual abelian gauge transformations).

The procedure is ill-defined only if the lowest two eigenvalues of the
covariant Laplacian coincide, or if $\bar\phi_x = 0$ for some $x$.
However, the set of such configurations has measure zero.  In practice,
we might get into problems when either the difference between the two lowest
eigenvalues, or $\phi_x$ for some $x$, is zero within the numerical 
precision of the computer, but, as it turns out, this never occurs.

The possibility of a zero in $\bar\phi$ (not necessarily at a lattice site)
is actually quite interesting.  It means that the gauge fixing is 
ill-defined at that point, which is precisely what identifies a
magnetic monopole in the abelian projection.
In fact, the 't~Hooft-Polyakov monopole (dyon) in the radial gauge satisfies
the continuum equivalent (see below) of the
LAG, with the solution for $\bar\phi$ 
equal to the Higgs field ($A_4$)
of this configuration (which has a zero at the origin)!

In the continuum limit, the stationarity conditions corresponding to 
the LAG are
\be
\sum_\mu \left( -\partial^2_\mu + (\bar A^1_\mu)^2 
    + (\bar A^2_\mu)^2 \right) \rho \= \lambda \, \rho ,
\label{cont1}
\ee
\be
\sum_\mu \left( \partial_\mu \,\mp\, i \bar A^3_\mu \right)
 \left(\rho^2 \bar A^\pm_\mu\right) \= 0 .
\label{cont2}
\ee
(Note that $\rho(x)$ depends on the gauge orbit as a whole only.)\@
This corresponds to minimization of the quantity
\be
\frac{\int_V \rho^2 \, \left[ (A^1_\mu)^2 + (A^2_\mu)^2 \right]}
     {\int_V \rho^2}
\label{rhominim1}
\ee
(its minimum value being the smallest eigenvalue $\lambda$ of the covariant
Laplacian)
or, on the lattice,
\be
\sum_{x,\mu} \left\{ 1 - \rho_x^{(\{U\})} \rho_{x+\hat\mu}^{(\{U\})\,}
  \half \Tr \left[\sigma_3 U_{\mu,x}^{(\Omega)}
            \sigma_3 U_{\mu,x}^{(\Omega)\,+} \right] \right\}
\! . \label{rhominim2}
\ee

Note that in all these formulas one recovers the corresponding
expression for the MAG \cite{smitvds} by setting $\rho$ equal to 1.
In fact, in the continuum limit $\beta\ra\infty$, the LAG-fixed configuration
is characterized by
$\bar U_{\mu,x}\ra \id$ and $\rho_x \ra 1$, and one sees that
LAG and MAG converge to each other.

At finite $\beta$, the function $\rho$ acts like a kind of ``measure of 
local smoothness'':
eq.~\rf{cont1} suggests that $\rho(x)$ will be small when
$(\bar A^1_\mu)^2(x) + (\bar A^2_\mu)^2(x)$
wants to be large.  
For example, in the centre of the 't~Hooft-Polyakov dyon
the non-abelian field components
$\bar A_\mu^{1,2}$ blow up while $\rho =0$.
Such divergences thus bear a relatively low penalty in the gauge fixing
functional \rf{rhominim1}:
the requirement of smoothness appears to tell the LAG to treat monopoles 
more mildly than the MAG does.

\section{NUMERICAL TESTS}
The following table contains some results for a $6^4$ lattice at 
various $\beta$ values.
\vspace*{5mm}

\centerline{
\begin{tabular}{|c||c|c|c|c|}
\hline
$\beta$       & $2.2     \!$ & $     2.4\!$ & $     2.6\!$ & $     2.8\!$ \\
\hline \hline
$\lambda_1$   & $2.123(3)\!$ & $1.872(8)\!$ & $1.677(2)\!$ & $1.529(2)\!$ \\
\hline
$\lambda_2$   & $2.154(3)\!$ & $1.926(8)\!$ & $1.759(2)\!$ & $1.631(3)\!$ \\
\hline
$\Delta\rho$  & $0.512(3)\!$ & $0.402(1)\!$ & $0.293(2)\!$ & $0.243(2)\!$ \\
\hline
\#$_L$        & $492(6)  \!$ & $149(13) \!$ & $28.2(8) \!$ & $6.6(4)  \!$ \\
\hline
\#$_M$        & $428(4)  \!$ & $103(10) \!$ & $16.8(5) \!$ & $4.2(3)  \!$ \\
\hline
\end{tabular}
\vspace*{5mm}
}
$\lambda_{1,2}$ are the two lowest eigenvalues of the covariant Laplacian, and
$\Delta\rho$ is the average fluctuation of $\rho$ over a configuration.  
For $\beta\ra\infty$, $\lambda_1$ tends to zero, $\lambda_2$ to one
(becoming 24-fold degenerate), and $\Delta\rho$ to zero.
However, one should keep in mind that
this limit is unphysical since it corresponds to zero physical volume.

\#$_L$ and \#$_M$ are the 
numbers of dual links transmitting monopole current, for LAG and MAG
respectively.
Note that the monopole numbers are somewhat higher in the LAG.  This
should {\em not\/} be interpreted as a sign that the LAG is ``worse''
than the MAG, as might be suggested by the empirical fact that ``bad''
gauges usually have very high monopole numbers associated to them.
Rather, it may be regarded as an indication of the number of monopoles
present in a situation of optimal smoothness.

Fig.~1 shows a comparison of the monopole locations as determined
by the LAG and the MAG, for a randomly chosen configuration at $\beta=2.4$.
It is interesting to see the large overlap, which is a sign of the 
similarity between the two gauges.
\begin{figure}[tb]
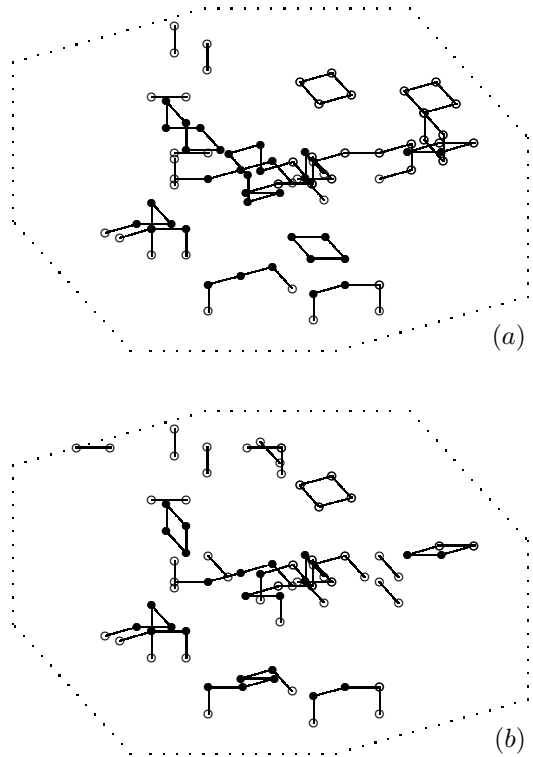

\vspace*{1mm}
\hspace*{-12.5mm}
\input fig_mon_lag
\hspace*{-12.5mm}
\input fig_mon_mag
\vspace*{-1.45cm}
\caption{Monopole loops for a $6^4$ configuration at $\beta=2.4$,
using LAG $(a)$ and MAG $(b)$.  
Open circles mark sites in the boundary of the (dual) lattice,
full circles indicate sites in the interior.}
\end{figure}

\end{document}